\documentclass{article}
\usepackage{spconf,amsmath,graphicx}
\usepackage{booktabs}
\usepackage{microtype}
\graphicspath{{./figures/}}

\usepackage{cite}

\title{Self Super-Resolution for Magnetic Resonance Images using~Deep~Networks}
\name{Can Zhao$^{1}$, Aaron Carass$^{1,2}$, Member, IEEE, Blake E.
Dewey$^{1}$, and Jerry~L.~Prince$^{1}$, Fellow, IEEE}

\address{$^{1}$ Dept. of Electrical and Computer Engineering, The
Johns Hopkins University,\\
Baltimore,~MD~21218~USA\\
$^2$Dept. of Computer Science, The Johns Hopkins University,\\
Baltimore, MD 21218 USA}

\begin{document}
%
\maketitle
\begin{abstract}
High resolution magnetic resonance~(MR) imaging~(MRI) is desirable in
many clinical applications; however, there is a trade-off between
resolution, speed of acquisition, and noise. It is common for MR
images to have worse through-plane resolution~(slice thickness) than
in-plane resolution. In these MRI images, high frequency information
in the through-plane direction is not acquired, and cannot be resolved
through interpolation. To address this issue, super-resolution methods
have been developed to enhance spatial resolution. As an ill-posed
problem, state-of-the-art super-resolution methods rely on the
presence of external/training atlases to learn the transform from low
resolution~(LR) images to high resolution~(HR) images. For several
reasons, such HR atlas images are often not available for MRI
sequences. This paper presents a self super-resolution~(SSR)
algorithm, which does not use any external atlas images, yet can still
resolve HR images only reliant on the acquired LR image. We use a
blurred version of the input image to create training data for a
state-of-the-art super-resolution deep network. The trained network is
applied to the original input image to estimate the HR image. Our SSR
result shows a significant improvement on through-plane resolution
compared to competing SSR methods.
\end{abstract}

\section{Introduction}
\label{s:intro}
The spatial resolution of magnetic resonance~(MR) images~(MRI) is
chosen based on imaging time, desired signal to noise ratio, and other
factors. Ultimately, the spatial resolution is limited by the amount
of $k$-space acquired in the Fourier domain. To facilitate faster, and
therefore cheaper, MRI acquisitions, it is common for MR images to have
worse through-plane resolution~(slice thickness) than in-plane
resolution. This means that the in-plane data has a relatively complete sampling
of $k$-space along an appropriate axis, whereas data in the
through-plane direction is bandlimited within the corresponding
$k$-space.

One common way to address the resolution mismatch between the in-plane
and through-plane directions is to upsample the data to an isotropic
resolution. This, however, results in images with partial volume
artifacts that lead to degraded image analysis in subsequent
processing. More appropriate approaches for estimating the high
frequency information are known as super-resolution~(SR) methods, as
they are meant to enhance the spatial resolution. SR has been a
well-explored technique in computer vision. Popular methods include
neighbor embedding regression~\cite{chang2004super,
timofte2014a+}, random forest
approaches~\cite{schulter2015fast, salvador2015naive}, and
state-of-the-art CNN methods such as those reported for the NTIRE 2017
Challenge~\cite{timofte2017ntire}, most of which were based on
VDSR~\cite{kim2016accurate} or SRResNet~\cite{ledig2016photo} as
a baseline model. EDSR~\cite{lim2017enhanced} which was modified 
from SRResNet~\cite{ledig2016photo}
had the best performance
in the NTIRE 2017 Challenge.

Unfortunately, all the NTIRE 2017 Challenge methods require external
paired atlas images to learn the transformation from low~(LR) to high
resolution~(HR). This is not a desirable situation for MR imaging, as
training data is not generally available because: 1)~scanner gain
means that even data acquired on the same scanner will have a
different dynamic range; 2)~acquiring HR MR data is difficult due to
scan times, patient motion and safety; and 3)~it is difficult to match
atlas and subject image resolutions perfectly. 
In contrast, exsiting single image self super-resolution~(SSR)
methods~\cite{huang2015cvpr,weigert2017isotropic,jog2016self} downsample the LR image to create a lower
resolution (LR$_2$) image and learn the mapping from LR$_2$ to LR; and
subsequently apply the mapping to LR with the goal of approximating
HR. 

In this paper, we build upon the work of Jog et al.~\cite{jog2016self}
which had an alternative approach to SSR. Jog et al. used the fact
that MR images are inherently anisotropic to learn a regression
between LR and HR images. The approach generated new additional
images, each of which is LR along a certain direction, but is HR in
the plane normal to it. Thus, each new image contributed information
to a new region of Fourier space. Then the collection of images is
combined using Fourier Burst Accumulation~\cite{delbracio2015tci}. We
modify the deep network framework of EDSR~\cite{lim2017enhanced},
while incorporating the ideas of Jog et al. to provide the training
data. Thus, we present a single image SSR deep network framework for MR
images. We refer to it as EDSSR.

\section{Method}
\label{s:method}
Consider an HR image $I(x,y,z)$ reconstructed from its $k$-space signal
$F(u, v, w)$. To save acquisition time and to improve signal-to-noise ratio, 
$F(u, v, w)$ is bandlimited
along the $w$-axis. The missing portion of $k$-space is filled with
$0$ and we refer to this Fourier space as $F_w(u, v, w)$, with the
reconstructed image denoted as $I_z(x, y, z)$. $I_z(x, y, z)$ has the same
digital resolution as $I(x, y, z)$, and has low spatial resolution in
the $z$ direction. The goal is to restore $I(x, y, z)$ from $I_z(x, y,
z)$ without any external training data.

\begin{figure}[!tb]
\centering
\begin{tabular}{c}
\includegraphics[width=8.5cm]{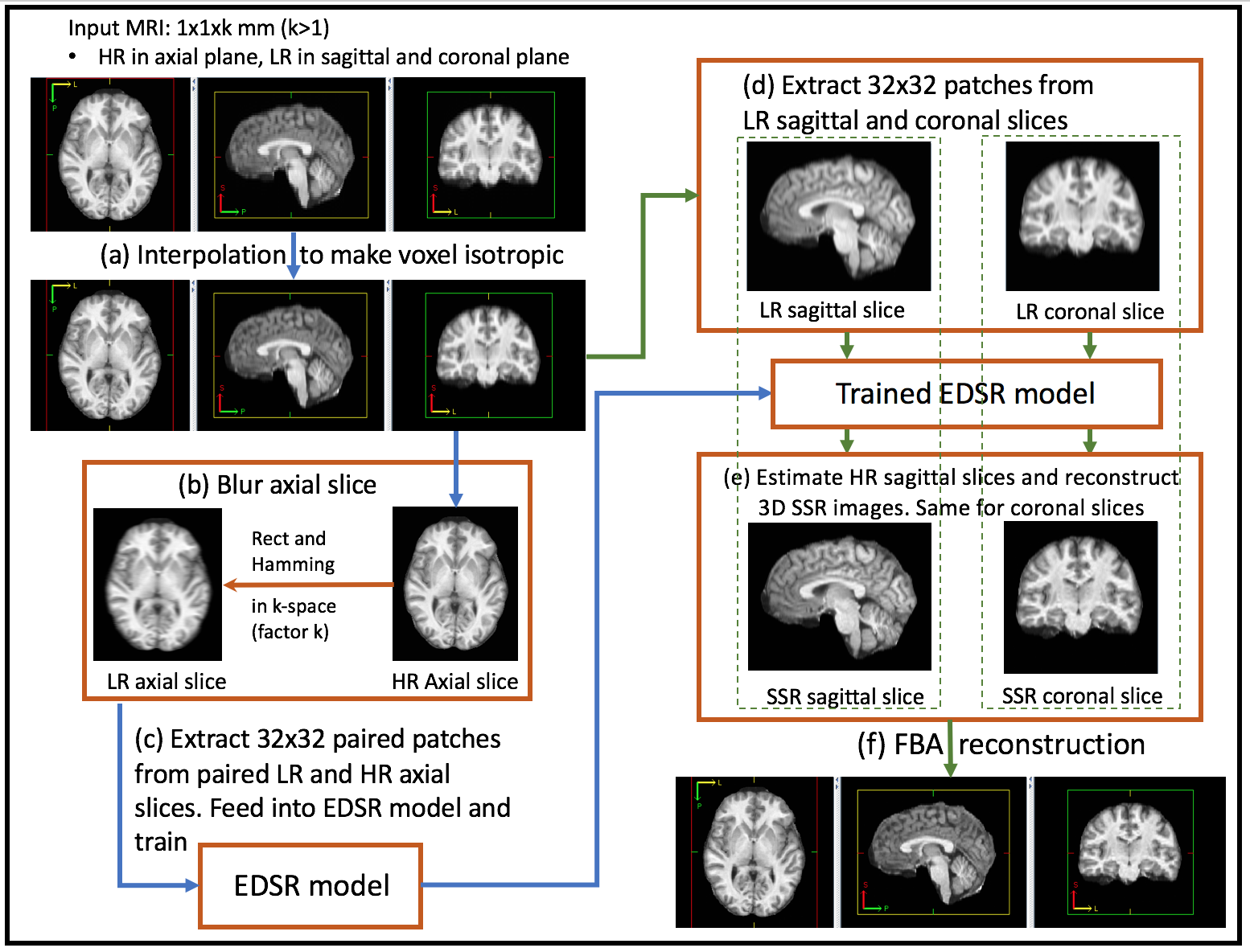}
\end{tabular}
\caption{The framework of our algorithm.}
\label{f:workflow}
\end{figure}

An overview of our framework is provided in Fig.~\ref{f:workflow}.
Given a LR input image $I_z(x, y, z)$ with digital resolution $1
\times 1 \times k$ where $k > 1$, we know that it has isotropic
digital resolution in the $xy$-plane, and low resolution in the
$z$-axis, i.e. its axial slices have high resolution $1 \times 1$,
while sagittal and coronal slices have low resolution $k \times 1$. We
first use interpolation to make the volume isotropic; we use
zero-padding in $k$-space though alternatives like cubic spline
interpolation~(BSP) may also be used. As the high frequency
information in $F_w(u, v, w)$ is completely missing, this
interpolation does not improve the spatial resolution. Instead, we
need a non-linear model to estimate the HR image from the LR image.
State-of-the-art SR models using deep networks
(EDSR~\cite{timofte2017ntire}) use paired HR and LR training data.
However, in our setting, no external training data is available.

To circumvent the lack of training data, we simulate training data
from the input image $I_z(x, y, z)$. To do this, we note that each 2D
axial slice $I_z(x, y)$ in $I_z(x, y, z)$ is actually an HR image,
while each coronal slice $I_z(x, z)$ and each sagittal slice $I_z(y,
z)$ are LR images. By blurring $I_z(x, y, z)$ along the $x$-axis, we
can obtain a LR image in both the $x$ and $z$ directions, which we
denote as $I_{zx}(x, y, z)$. The blurred image $I_{zx}(x, y, z)$ and
input image $I_z(x, y, z)$ are used as our training data. The axial
slices $I_{zx}(x, y)$ of $I_{zx}(x, y, z)$ have resolution of $k
\times 1$, while the axial slices $I_{z}(x, y)$ of $I_z(x, y, z)$ have
resolution of $1\times1$. Now, if we can learn a mapping from LR image
$I_{zx}(x,y)$ to HR image $I_{z}(x,y)$, we can apply the mapping to LR
images $I_z(x, z)$ and $I_z(y, z)$ to estimate the HR images $I(x, z)$
and $I(y, z)$.

We use EDSR, a state-of-the-art deep network SR model, to learn the
transformation. After applying the trained model to coronal slices,
$I_z(x, z)$, we get $\hat{I}^{y}(x, z)$, an estimate of $I(x, z)$.
By stacking together each $\hat{I}^{y}(x, z)$, we have $\hat{I}^{y}(x,
y, z)$. We can repeat this process for $I_z(y, z)$ to generate
$\hat{I}^{x}(y, z)$ and subsequently $\hat{I}^{x}(x,y,z)$. Similar to
Jog et al.~\cite{jog2016self}, we use Fourier Burst
Accumulation~(FBA)~\cite{delbracio2015tci} to reconstruct
$\hat{I}(x,y,z)$ from the two images $\hat{I}^{y}(x, y, z)$ and
$\hat{I}^{x}(x,y,z)$. Complete details of the construction of the
training data and our modifications to EDSR are listed below.

\textbf{Training data extraction:}~We first blur $I_z(x, y, z)$ in the
$x$-axis and obtain $I_{zx}(x, y, z)$. To simulate the data
acquisition process in MRI, we use the low pass filter on the
$k$-space signal $F_w(u, v, w)$. A rect function on the $u$-axis is
multiplied with $F_w(u, v, w)$, generating $F_{wu}(u, v, w)$ while
guaranteeing no high frequency information on the $u$-axis. For 3D
acquired MR images, a window function might be applied along the
$w$-axis during reconstruction to avoid ringing. 
To increase the amount of available training
data, we use rotated versions of the original image. If we rotate
$I_{z}(x, y, z)$ in the $xy$-plane, the rotated images $R_{xy}(\theta)
\circ I_{z}(x, y, z)$ still has resolution $1 \times 1 \times k$. We
can therefore do the same blurring to the rotated images
$R_{xy}(\theta) \circ I_{z}(x, y, z)$ to obtain more training data. We
randomly extract $32 \times 32$ patches in each slice $z$ of
$I_{zx}(x, y, z)$ and it matching pair in $I_z(x, y, z)$, as well as
from the rotated images; feeding these paired 2D HR and LR patches
into the deep network to train it.

\textbf{EDSR model:}~We use the default number of layers and loss 
as in the original EDSR~\cite{timofte2017ntire} except for one modification.
The original EDSR~\cite{timofte2017ntire}
framework used an upsampling layer in the model to make the image
pixels isotropic. We in contrast upsample the LR image to an isotropic
resolution prior to being input into the deep network, thus there is
no upsampling layer in the model. A reason for this, is the upsampling
factor $k$ varies among different data sets. By having this factor be
part of the network would force the model to be changed and thus
retrained for each new data set. By having a fixed network structure
and doing the isotropic resampling outside the EDSR framework, allows
the network weights to be reused between data sets with only fine-tune
of the pre-trained network weights. This modification can cut the time
cost by approximately $50$-$90\%$ depending on the similarity between the current
data set and the pre-trained data set. An additional benefit being
that EDSR required the factor $k$ to be an integer which is generally
not true of MR images, however our framework allows $k$ to be a non-integer value.

\section{Experiments}
\label{sec:exp1}
\begin{figure}[t]
\begin{minipage}[b]{.49\linewidth} \centering
\centerline{\includegraphics[height=4.2cm]{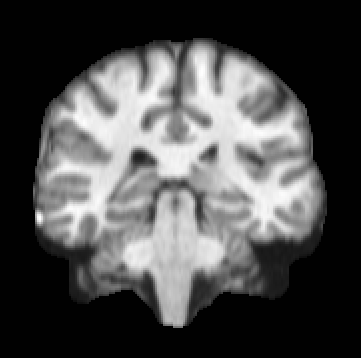}}
 \centerline{\textbf{(a)}~BSP ($k=2$)}\medskip \end{minipage} \hfill
 \begin{minipage}[b]{0.49\linewidth} \centering
 \centerline{\includegraphics[height=4.2cm]{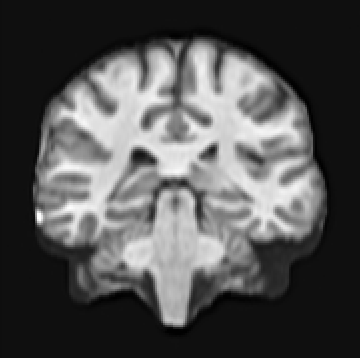}}
 \centerline{\textbf{(b)}~SSR ($k=2$) \cite{jog2016self}}\medskip
 \end{minipage}

\begin{minipage}[b]{.49\linewidth} \centering
\centerline{\includegraphics[height=4.2cm]{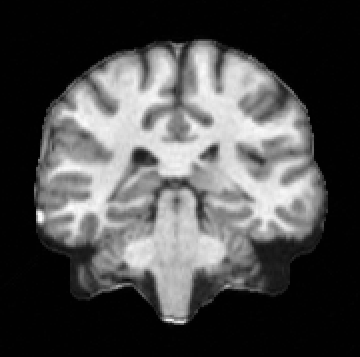}}
 \centerline{\textbf{(c)}~EDSSR ($k=2$)}\medskip \end{minipage} \hfill
 \begin{minipage}[b]{0.49\linewidth} \centering
 \centerline{\includegraphics[height=4.2cm]{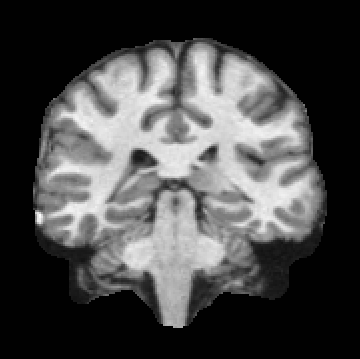}}
 \centerline{\textbf{(d)}~HR (1mm)}\medskip \end{minipage}
\caption{Coronal views of the \textbf{(a)}~cubic B-spline (BSP)
interpolated image of the 2 mm LR image,
\textbf{(b)}~result using SSR~\cite{jog2016self},
\textbf{(c)}~our result EDSSR using EDSR and FBA, 
\textbf{(d)}~HR ground truth image}
\label{fig:exp1-2}
\end{figure}

\begin{figure}[t]
\begin{minipage}[b]{.49\linewidth}
 \centering
 \centerline{\includegraphics[height=4.2cm]{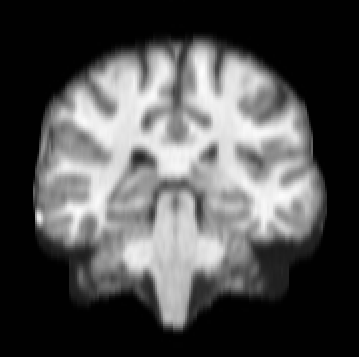}}
 \centerline{\textbf{(a)}~BSP ($k=3$)}\medskip
\end{minipage}
\hfill
\begin{minipage}[b]{0.49\linewidth}
 \centering
 \centerline{\includegraphics[height=4.2cm]{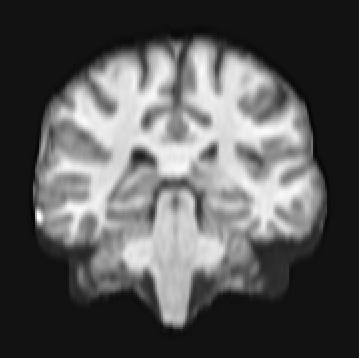}}
 \centerline{\textbf{(b)}~SSR ($k=3$) \cite{jog2016self}}\medskip
\end{minipage}

\begin{minipage}[b]{.49\linewidth} \centering
\centerline{\includegraphics[height=4.2cm]{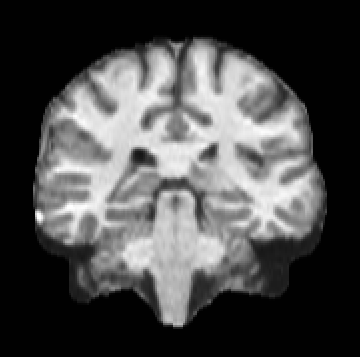}}
 \centerline{\textbf{(c)}~EDSSR ($k=3$)}\medskip \end{minipage}
 \hfill \begin{minipage}[b]{0.49\linewidth} \centering
 \centerline{\includegraphics[height=4.2cm]{2mm_grth}}
 \centerline{\textbf{(d)}~HR (1mm)}\medskip \end{minipage}
\caption{Coronal views of the \textbf{(a)}~cubic B-spline (BSP)
interpolated image of the 3 mm LR image,
\textbf{(b)}~result using SSR~\cite{jog2016self},
\textbf{(c)}~our result EDSSR using EDSR and FBA, 
\textbf{(d)}~HR ground truth image}
\label{fig:exp1-3}
\end{figure}

We ran our algorithm on $T_1$-weighted Magnetization Prepared Rapid
Gradient Echo~(MPRAGE) images from 20 subjects of the
Neuromorphometrics dataset, and compare with other methods. The ground
truth HR image has a resolution of $1 \times 1 \times 1$~mm. As the
input of SSR algorithms, the LR images are simulated by
downsampling in the $z$-axis from the ground-truth HR image by a
factor $k$ of 2 and 3. Cubic B-spline interpolation~(BSP) and
SSR~\cite{jog2016self}, as well as our method are shown for the
downsampling factors of 2 and 3 in Fig.~\ref{fig:exp1-2} and
Fig.~\ref{fig:exp1-3}, respectively. Visually, our EDSSR approach
can significantly improve through-plane resolution compared to BSP and
SSR~\cite{jog2016self}, especially in Fig.~\ref{fig:exp1-3} when the
downsampling scale factor is 3. A comparison of the PSNR values for
the three methods is shown in Table~\ref{t:ex1}, with the SSIM for all
three methods being shown in Table~\ref{t:ex2}. The ‘*’ indicates that the
results are statistically significantly greater than all the methods
using a one-tailed t-test and Wilcoxon rank-sum tests. 
Our EDSSR approach shows superiority in SSIM and PSNR 
for input LR images with resolution of $1 \times 1 \times 3$mm.

\begin{table}[!tb]
\caption{Mean PSNR values~(dB) for 20 subjects. The ‘*’ indicates that
the results are statistically significantly greater than all the
methods using a one-tailed t-test and Wilcoxon rank-sum tests for the
PSNR metric.}
\label{t:ex1}
\centering
\begin{tabular}{lc lc lc l} 
\toprule
\textbf{LR} && \textbf{BSP} && \textbf{SSR} && \textbf{EDSSR}\\
\cmidrule{1-7}
\textbf{2mm} && 35.99 && 37.98* && 35.14 \\
\textbf{3mm} && 31.98 && 33.49 && 34.44*\\
\bottomrule
\end{tabular}
\end{table}

\begin{table}[!tb]
\caption{Mean SSIM values for 20 subjects. The ‘*’ indicates that the
results are statistically significantly greater than all the methods
using a one-tailed t-test and Wilcoxon rank-sum tests for the SSIM
metric.}
\label{t:ex2}
\centering
\begin{tabular}{lc lc lc l} 
\toprule
\textbf{LR} && \textbf{BSP} && \textbf{SSR} && \textbf{EDSSR}\\
\cmidrule{1-7}
\textbf{2mm} && 0.9873* && 0.9832 && 0.9763\\
\textbf{3mm} && 0.9635 && 0.9678 && 0.9773*\\
\bottomrule
\end{tabular}
\end{table}


\begin{figure}[htb]
\begin{minipage}[b]{0.32\linewidth} \centering
\centerline{\includegraphics[height=2.05cm]{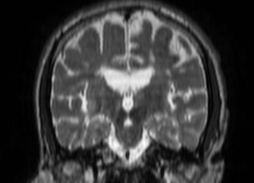}}
 \centerline{\textbf{(a)} BSP}\medskip \end{minipage} \hfill
 \begin{minipage}[b]{0.32\linewidth} \centering
 \centerline{\includegraphics[height=2.05cm]{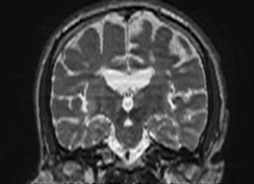}}
 \centerline{\textbf{(b)} SSR \cite{jog2016self}}\medskip
 \end{minipage} \hfill \begin{minipage}[b]{0.32\linewidth} \centering
 \centerline{\includegraphics[height=2.05cm]{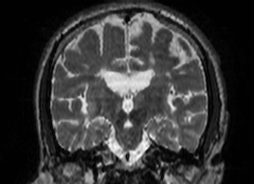}}
 \centerline{\textbf{(c)} EDSSR}\medskip \end{minipage}
\begin{minipage}[b]{.32\linewidth} \centering
\centerline{\includegraphics[height=2.25cm]{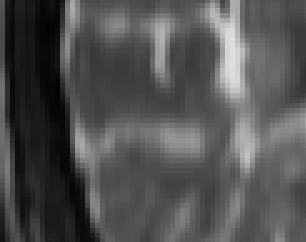}}
 \centerline{\textbf{(d)} BSP}\medskip \end{minipage} \hfill
 \begin{minipage}[b]{0.32\linewidth} \centering
 \centerline{\includegraphics[height=2.25cm]{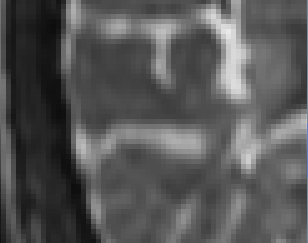}}
 \centerline{\textbf{(e)} SSR\cite{jog2016self}}\medskip
 \end{minipage} \hfill \begin{minipage}[b]{0.32\linewidth} \centering
 \centerline{\includegraphics[height=2.25cm]{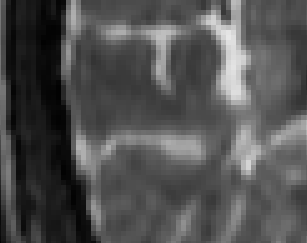}}
 \centerline{\textbf{(f)} EDSSR}\medskip \end{minipage}
\caption{Coronal views of the \textbf{(a)}~cubic B-spline (BSP)
interpolated image of the $0.83 \times 0.83 \times 2.20 $ mm LR image,
\textbf{(b)}~result using ANR and FBA~\cite{jog2016self},
\textbf{(c)}~our result using EDSR and FBA; zoomed images of the
\textbf{(d)}~BSP interpolated image, \textbf{(e)}~result using ANR and
FBA~\cite{jog2016self}, \textbf{(f)}~our result using EDSR and FBA,}
\label{fig:exp2}
\end{figure}

We also experimented using real LR image data. We used T2-weighted MRI 
acquired at $1.14 \times 1.14 \times 2.20$~mm
in 2D, and reconstructed at $0.83 \times 0.83 \times 2.20$~mm, with
the scale factor being $k = 2.20/0.83 \approx 2.65$. The results for
our method and SSR are shown in Fig.~\ref{fig:exp2}. The lower row
shows a zoom on an edge. Although there is no ground truth, visually
our approach gives significantly better results than BSP and
SSR~\cite{jog2016self}.

\section{Conclusion and Discussion}
\label{s:typestyle}
This paper presents an SSR approach which does not need any
external training data to estimate an HR image from a LR image. It uses
patches in blurred axial slices of the image itself to create paired
training data, and the state-of-the-art SR model from EDSR to model
the mapping from LR patches to HR patches. Finally, it uses
FBA~\cite{delbracio2015tci} to reconstruct the final image, taking
advantage of SR results from different orientations. The results are
significantly better than competing methods, in particular with
decreasing through-plane resolution (ie. increasing slice thickness).

Deep networks are known to be dependent on ``good'' atlas data. The
variance between atlas and subject could cause severe over-fitting.
However, real-world MR images have relatively varying intensities.
This SSR algorithm not only fits in the clinical practice that the
MR images with high through-plane resolution are usually unavailable,
but also guarantees the intensity invariance between atlas and
subject, which reduces the over-fitting of the deep network. In
summary, this SR algorithm can improve image resolution without any
external data.

\bibliographystyle{IEEEbib}
\bibliography{refs}

\end{document}